\documentclass[pdflatex,sn-mathphys-num]{sn-jnl}

\usepackage{graphicx}%
\usepackage{multirow}%
\usepackage{amsmath,amssymb,amsfonts}%
\usepackage{amsthm}%
\usepackage{mathrsfs}%
\usepackage[title]{appendix}%
\usepackage{xcolor}%
\usepackage{textcomp}%
\usepackage{manyfoot}%
\usepackage{booktabs}%
\usepackage{algorithm}%
\usepackage{algorithmicx}%
\usepackage{algpseudocode}%
\usepackage{listings}%
\usepackage{threeparttable}%

\usepackage[left]{lineno}

\newcommand{\kms}{km\,s$^{-1}$}
\def\msun{{\rm\,M_\odot}}

\def\ie{{ i.e.,\ }}

\begin{document}

\title[C-19]{The primordial nature of the C-19 stellar stream}


\author*[1]{\fnm{Kim A.} \sur{Venn}}\email{kvenn@uvic.ca}

\author*[2,3,4]{\fnm{Zhen} \sur{Yuan}}
\email{zhen.yuan@nju.edu.cn}

\author[4,5]{\fnm{Nicolas F.}\sur{Martin}}

\author[]{\fnm{Anya}\sur{Dovgal}}

\author[]{\fnm{Daria}\sur{Zaremba}}

\author[]{\fnm{Else}\sur{Starkenburg}}


\author[]{\fnm{Felipe} \sur{Gran}}

\author[]{\fnm{Christian R.} \sur{Hayes}} 

\author[]{\fnm{Vanessa} \sur{Hill}}

\author[]{\fnm{Chiaki}\sur{Kobayashi}}
 
\author[]{\fnm{Carmela}\sur{Lardo}}

\author[]{\fnm{Alan W.} \sur{McConnachie}} 

\author[]{\fnm{Tadafumi} \sur{Matsuno}}

\author[]{\fnm{Martin} \sur{Montelius}}

\author[]{\fnm{Vinicius} \sur{Placco}}

\author[]{\fnm{Federico} \sur{Sestito}}


\author[]{\fnm{Anke} \sur{Ardern-Arentsen}} 

\author[]{\fnm{Guiseppina} \sur{Battaglia}} 

\author[]{\fnm{Piercarlo} \sur{Bonifacio}} 

\author[]{\fnm{Raymond} \sur{Carlberg}}

\author[]{\fnm{Sebastien} \sur{Fabbro}}

\author[]{\fnm{Morgan} \sur{Fouesneau}}

\author[]{\fnm{Rodrigo} \sur{Ibata}}

\author[]{\fnm{Pascale} \sur{Jablonka}}

\author[]{\fnm{Jaclyn} \sur{Jensen}}

\author[]{\fnm{Georges} \sur{Kordopatis}}

\author[]{\fnm{Madelyn } \sur{McKenzie}}

\author[]{\fnm{Julio F.} \sur{Navarro}} 

\author[]{\fnm{John S.} \sur{Pazder}}

\author[]{\fnm{Ruben} \sur{Sanchez-Janssen}} 

\author[]{\fnm{Simon T. E.} \sur{Smith }}

\author[]{\fnm{Akshara} \sur{Viswanathan}}

\author[]{\fnm{Sara} \sur{Vitali}}

\author[]{\fnm{Long} \sur{Wang}} 

\author[]{\fnm{Zhen} \sur{Wang}} 


\affil[1]{\orgdiv{Department of Physics and Astronomy}, \orgname{University of Victoria}, \orgaddress{\street{3800 Finnerty Road}, \city{Victoria}, \postcode{V8P 5C2}, \state{BC}, \country{Canada}}}

\affil[2]{\orgdiv{School of Astronomy and Space Science}, \orgname{Nanjing University}, \orgaddress{\city{Nanjing}, \postcode{210093}, \state{Jiangsu}, \country{China}}}

\affil[3]{\orgdiv{Key Laboratory of Modern Astronomy and Astrophysics}, \orgname{Nanjing University, Ministry of Education}, \orgaddress{\city{Nanjing}, \postcode{210093}, \state{Jiangsu}, \country{China}}}

\affil[4]{\orgdiv{Observatoire Astronomique de Strasbourg}, \orgname{CNRS, Universit\'e de Strasbourg}, \orgaddress{\street{UMR 7550}, \city{Strasbourg}, \postcode{F-67000}, 
\country{France}}}

\affil[5]{\orgname{Max-Planck-Institut f\"{u}r Astronomie}, \orgaddress{\street{K\"{o}nigstuhl 17}, \city{Heidelberg}, \postcode{D-69117}, \country{Germany}}}










\abstract{  

Stellar streams, remnants of compact star systems stretched out by the tidal forces of the Milky Way, offer a unique way to study stellar populations that formed billions of years ago
\citep{Bonaca2025NewA}. 
A particularly unique stream is C-19, the most metal-poor stellar stream known at less than a thousandth of the Sun's metallicity \citep{MartinVenn2022}. 
The nature of C-19 is not yet clear, with properties that resemble both star clusters and ultra faint dwarf galaxies, yet in either case its extremely low metallicity indicates very early star formation, 
$\lesssim 1$\,Gyr after the Big Bang \citep{El-Badry2018}. 
Here, we present the first detailed study on the nature of C-19 based on the chemical abundances of 14 member stars from high-resolution spectroscopy \cite{Venn2025}.
These reveal that C-19 formed stars in an early, rapid, and prolific star formation event, with mild inhomogeneous mixing of elements produced in massive stars. There is otherwise no evidence for subsequent star formation, multiple stellar populations, nor chemical evolution.
Although C-19 is currently disrupted in the Milky Way halo, it offers a rare and complementary window into the details of star formation and chemical evolution in the early universe, ideal for comparisons with current studies of primordial star formation in the high-redshift universe \citep{Curti2024JWST}.
}


\maketitle

\begin{figure}[]
\centering
\includegraphics[width=0.9\textwidth]{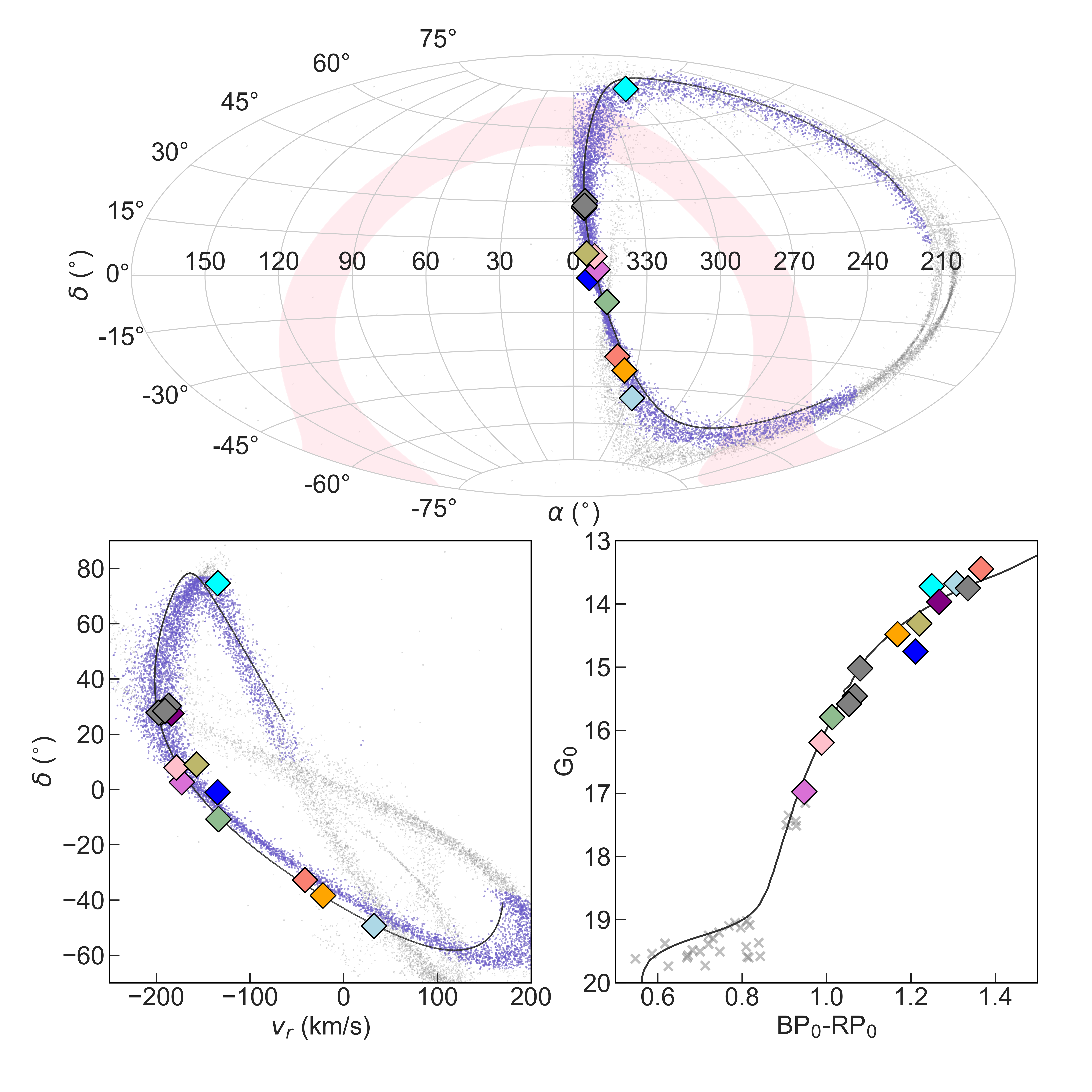}
\caption{The upper panel shows fourteen C-19 members from Table~\ref{tab:targets} in the celestial coordinates ($\alpha$, $\delta$), where the transparent pink band represents the disk region with $|b|< 10^{\circ}$ in Galactic coordinates. The eight members in the outer stream are colored in the same way as Fig.~\ref{fig:srba}, one of which is separated by the disk in the North. The lower left panel shows the velocities of the C-19 members along $\delta$. The particles from the simulation (see Metho are plotted as gray dots, and those close to the observed C-19 stream is highlighted in purple, following the stream orbit, shown as solid black line. The lower right panel shows the color-magnitude diagram of the C-19 members as well as the candidates without radial velocity information. The solid black line here represents a 12-Gyr isochrone with [Fe/H]$=-2.2$ at the distance of 18 kpc, from the standard PARSEC model \citep{bressan2012}.}
\label{fig:orbit}
\end{figure}

Studies of the stellar populations in Milky Way (MW) streams help to identify the characteristics of their progenitors while also being useful in deciphering the accretion history and dark matter profile of the MW.
A particularly unique system is C-19, the most metal-poor stellar stream known ([Fe/H] $=-3.4\pm0.20$)\footnote{Chemical abundances are provided using the standard notation, [X/Y] = log (X/Y) - log (X/Y)$_\odot$, where X and Y are number densities.  Thus, [Fe/H]$=-3$ is equal to 1/1000th the iron metallicity of the Sun.}, where its $\gtrsim 10^4$ M$_\odot$ of stars are spread over 100 degrees across the sky \citep{MartinVenn2022, Yuan2025}. 
Using the new high-resolution ($R\simeq50,000$) GHOST spectrograph at the Gemini South observatory \cite{McC2024GHOST}, we have homogeneously analysed high signal-to-noise ($\ge 40$ per pixel) spectra for 8 individual stars across the extent of the C-19 stream \cite{Venn2025}. 
Target selection,  data analysis, and stellar parameters are summarized in Methods and the abundances for a subset of the 16 measured chemical elements are presented in Table~\ref{tab:params}. 
In addition, these detailed chemical abundances were combined with literature values for another 6 members with high-resolution, high signal-to-noise spectroscopy.
The resulting full sample of 14 member stars is used here to provide critical insight into the origin of C-19. The on-sky distribution of these stars, as well as their velocity and color-magnitude information are summarized in Figure~\ref{fig:orbit}. 
 
\begin{figure*}
    \setlength{\unitlength}{1cm}
    \includegraphics[width=13cm] {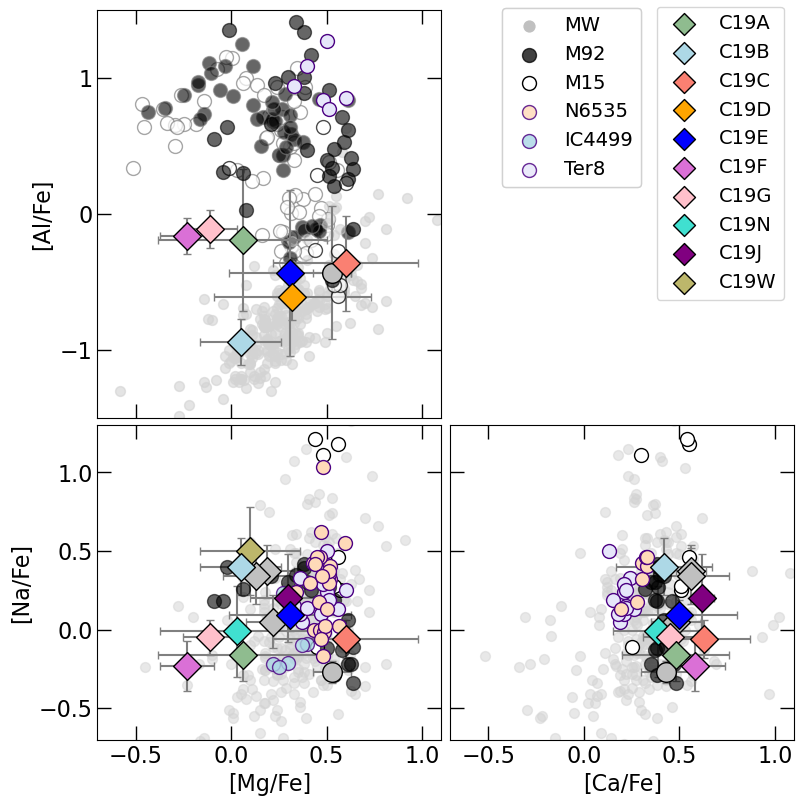} 
\caption{
Light element NLTE abundances for C-19 compared to metal-poor ([Fe/H]$<-2.5$) stars in the Milky Way and five globular clusters: 
two metal-poor GCs (M15/white, M92/darkgrey), and three low mass GCs (Ter 8/lavender, IC4499/lightblue, NGC6535/peach, and each with purple edges).
%
%
Three stars in the core of C-19 (C19KLM) are from published analysis of Gemini/GRACES spectra (grey diamonds).
%
References are in Section~\ref{secA1}.
Our measurements for the standard star HD122563 are shown as a large filled grey circle with black edges and errorbar.
    } 
\label{fig:chem}
\end{figure*}

This dataset confirms the very low metallicity of C-19 ($\langle$[Fe/H]$\rangle=-3.31^{+0.07}_{-0.08}$) and provides a strict constraint on its metallicity dispersion $\sigma_{\rm [Fe/H]}<0.18$ dex at the 95\% confidence level (see Figure~\ref{fig:FeH_contours} in Methods). 
Such a low iron metallicity dispersion is usually interpreted as a product of very rapid star formation during the birth of a star cluster, reaching a specific iron metallicity without enrichment from SN Ia and that depends on properties of the initial gas cloud \citep{Bastian2018, Bekki2019}. 
Figure~\ref{fig:chem} shows that the detailed abundances of the light elements (sodium Na, magnesium Mg, and aluminum Al) show narrow dispersions.
Al has a spread $\sigma_{\rm Al}\sim0.3$\,dex, while Mg and Na show only marginally resolved small spreads ($\sim0.1$\,dex), and Ca shows no significant dispersion (these dispersions are compared in Figure~\ref{fig:dispersions} in Methods).
The C-19 stars have very low Al 
when compared to the metal-poor MW globular clusters (i.e., M15 and M92, shown in Figure~\ref{fig:chem}),
as Al production is metallicity dependent \citep{Kobayashi2020, Pancino2017}.  Furthermore,  the MW globular clusters 
have large fractions of second generation stars (up to 50\%, \cite{Milone2020}) 
that are even further enhanced in Al through hot proton burning in the first generation of massive stars. Thus, the second generation stars, which form from recycled gas retained at the bottom of a cluster's potential, show an anti-correlation in Al with Mg produced by the Mg–Al proton-capture cycle \cite{Bastian2018, Vaca2024}. 
In contrast, C-19 does not show this pattern.  While Al and Mg show \textit{dispersions}, there are no (anti-)correlations, nor any relationships in Na, i.e., proton-capture reactions that lead to the Mg-Al cycle  will also trigger the Na-O cycle, as seen in Milky Way globular clusters.  Without observational evidence from these elements, C-19 seems to lack a second generation of stars. 

Lower-mass clusters, with stellar masses in the range $10^4$--$10^5\msun$ may be more similar to C-19, whose progenitor stellar mass is estimated to be 4--5$\times10^4\msun$ \citep{Yuan2025}. For these clusters, the presence of second generation stars is less clear, often inferred from little more than a minor spread in Na and no other element dispersions.
In Figure~\ref{fig:chem}, a dispersion in Na is seen in 3 low-mass globular clusters, with no spreads or anti-correlations with Mg or Al.
While their spread in Na is similar to that in C-19, their lack of an observed dispersion in Al and Mg is not similar to C-19.
The combination of minor light-element variations with no clear (anti)correlations and remarkably uniform iron abundances supports that C-19 formed rapidly in a gas cloud with some light inhomogeneous mixing of elements from its massive stars, as discussed further below.



\begin{figure*}
\begin{center}
    \includegraphics[width=11cm]{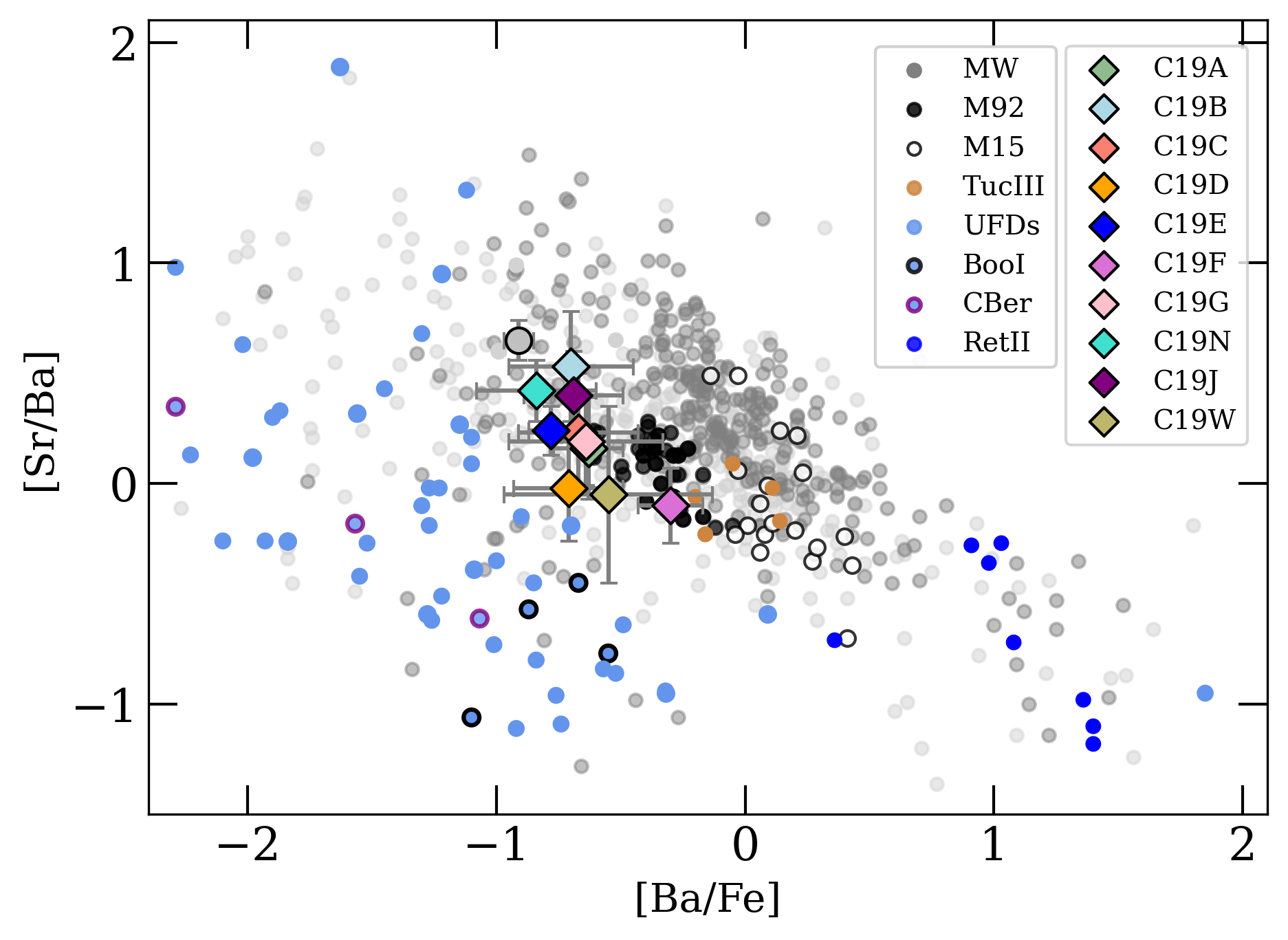}
    \includegraphics[width=11cm]{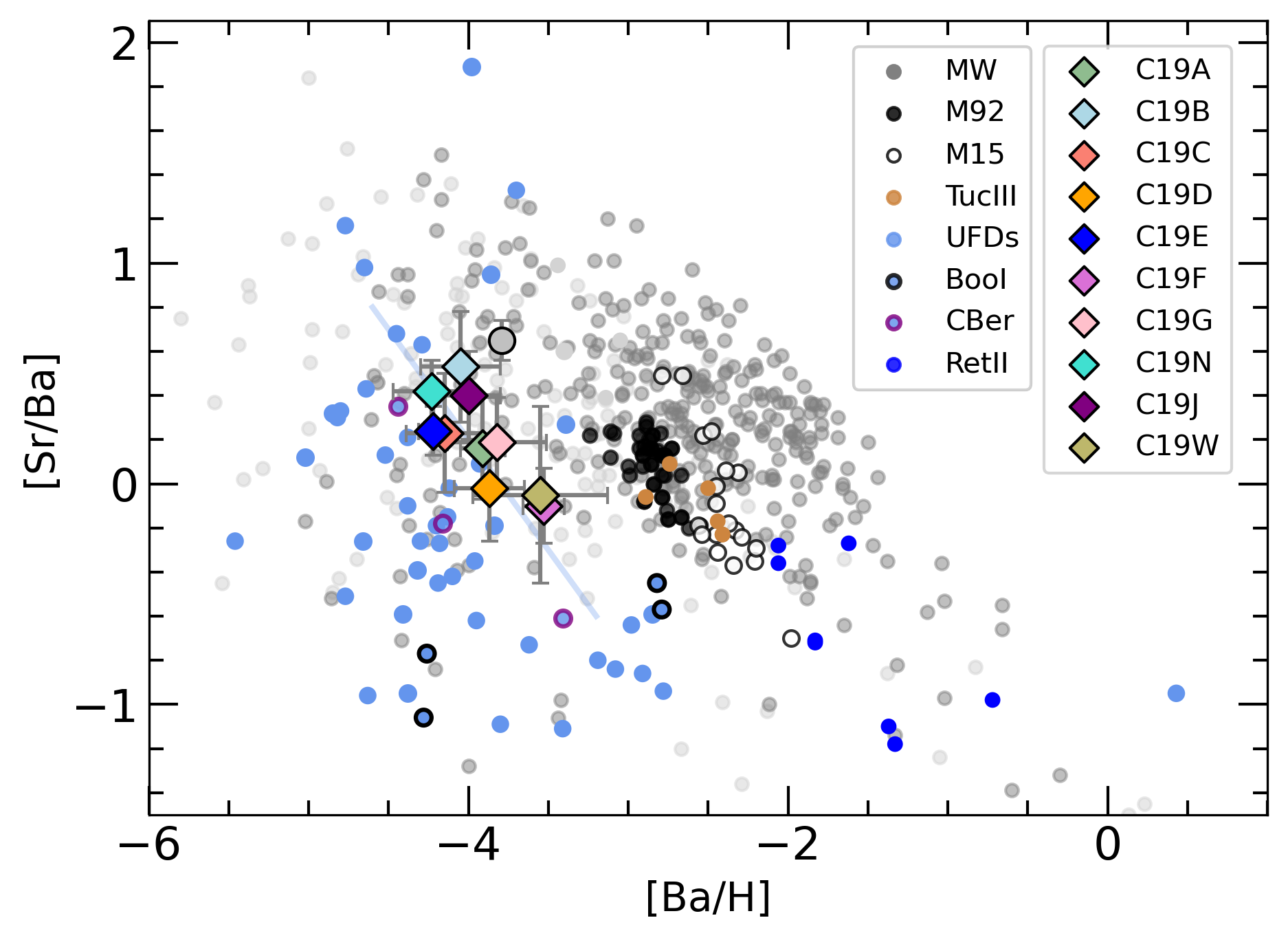}

\caption{
Sr and Ba NLTE abundances plotted as a function of [Ba/Fe] (top panel) and [Ba/H] (bottom panel, which includes a line of fixed [Sr/H] $=-3.8$ in light blue).
C-19 stellar abundances are compared to metal-poor stars in the Milky Way (SAGA \cite{Suda2017} in light grey, and \cite{HLi2022} in medium grey),
and the two metal-poor GCs M15 \cite[white;][]{Sobeck11, Garcia2024} and M92 \cite[black;][]{Kirby2023}.
Abundances from individual stars in UFDs are also shown, collected from the SAGA database (\cite{Suda2017}, references in Methods) and the GHOULS survey (\citep{Dovgal2025}, medium blue). 
Three UFDs with $>3$ stars with Sr and Ba each are highlighted: Bootes I (black edges from SAGA), Com Ber (purple edges from \cite{Sitnova2021}), and Ret II (bright blue from SAGA) which is offset due to a rare r-process event \cite{Ji2023Ret2}. We note that the system Tuc III (brown) resembles the metal-poor globular clusters more closely than UFDs.
References are in Section~\ref{secA1}.
Our measurements for the standard star HD122563 are shown as a large filled grey circle with errorbar.
%
    } 
\label{fig:srba}
\end{center}
\end{figure*}

Turning to heavier elements, Figure~\ref{fig:srba} compares the Ba and Sr abundances in C-19 to those measured in low metallicity stars in the Milky Way halo and UFDs.   
The Sr and Ba abundances in C-19 are typical for stars of their metallicity. The small dispersions in Sr and Ba resemble those of MW globular clusters, rather than the larger dispersions seen in UFDs (e.g., Bo\"otes~I, Coma Berenices, Reticulum II). 
In UFDs, such spreads are assumed to result from two channels: (1) chemical evolution, where ongoing star formation occurs over extended periods of time, enriching the interstellar gas with products from previous generations, and (2) stochastic and inefficient star formation that leads to fewer massive stars and significant inhomogeneous mixing, which produce a wide range of metallicities tied to the location of individual star formation events \cite{Revaz2018} and lower heavy element abundances overall \cite{Ji2019}.
%
%
The small Ba dispersion is consistent with the small Na and Mg dispersions discussed above, and together these are consistent with observations 
seen in \textit{first} population stars in the globular clusters M15 and M92.  Those have been interpreted as inhomogeneous mixing from the massive stars in the initial star forming event, including yields from an r-process site in massive stars with a short time delay \cite{Kirby2023, Henderson2025}.
The relatively high [Sr/Ba] ratio in C-19 favours yields from the weak r-process \cite{Sneden2008} or weak s-process \cite{Limongi2018}.
Incidentally, the overlap of C-19 with the metal-poor end of the MW halo means that a significant fraction of the extremely metal-poor MW halo is at least partly built out of stars like those found in C-19.

\begin{figure}[]
\hspace*{+0.2cm}
\includegraphics[width=1\textwidth]{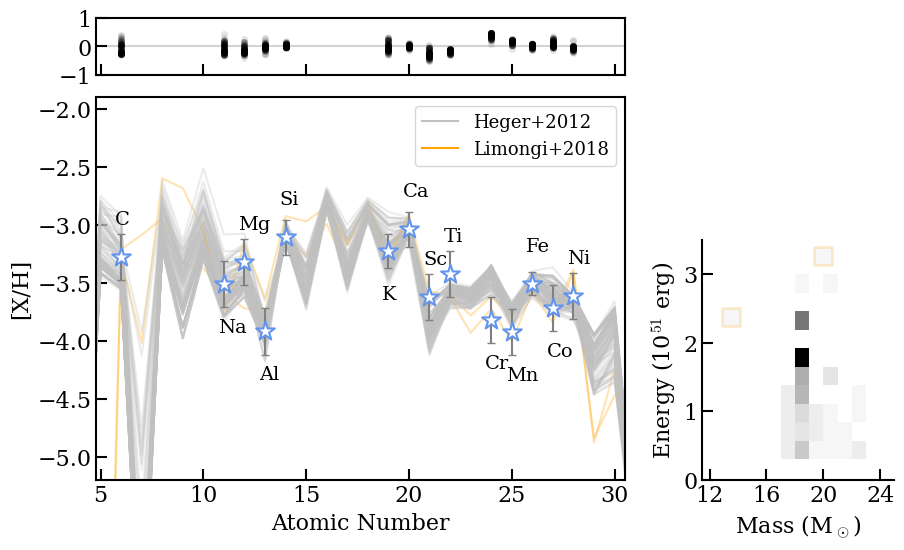}

\caption{The mean chemistry of C-19 \cite{Venn2025} (blue stars) compared to Pop III core-collapse supernova yields \cite{Heger2012} (grey) and two Pop II rapidly rotating massive star models \cite{Limongi2018} (orange).  All models are within 3$\sigma$ of the best fitting model (with $\chi^2$=0.47).  Residuals for each element between the observed and predicted abundance for each model are shown in the top plot.  
Right side shows the range in these best fit models in Mass and Explosion Energy, and where the two rapidly rotating massive star models (v$_{\rm rot}=300$ km/s) are shown with orange boxes.  
The general conclusion is that normal CCSN of stars near 20 M$_\odot$ are sufficient to explain the available chemical abundances in C19.  }
\label{fig:StarFit4}
\end{figure}

The overall chemistry of C-19 suggests that its progenitor must have formed stars rapidly (narrow metallicity range; no signatures of chemical evolution or multiple stellar generations) and early  (extremely low Fe and Al; chemical signatures only from massive stars).
At the low metallicity of C-19 ([Fe/H]$\simeq-$3.3), it is reasonable to ask if it formed out of gas enriched only in Population III stars, i.e., those with large masses and zero metallicity.
To test this hypothesis, we compare the average abundances in C-19 to Population III supernova yields \cite{Heger2012} in Figure~\ref{fig:StarFit4}. 
There are a large number of individual Pop III core collapse supernova (CCSN) models that can fit the measured abundances in C-19 to within 3$\sigma$ of the best fit model, and an even larger number of combinations of these models. However, we note that all of the models come from a small range in mass and explosion energy, i.e., 
a progenitor mass of $\approx 20\msun$, a typical explosion energy B $= 1-3\times10^{51}$\,erg, light mixing (mixing fraction, $f\sim0.1$), and standard dilution factors\footnote{We note that the fits do not include the r-process elements, as no elements heavier than the iron-group are predicted to form in CCSN of Pop III stars -- however there are other sites for the main or weak  r-process in massive stars not included in these yields.}.
It is reasonable therefore to consider that C-19 may have even been enriched by a single CCSN zero-metallicity massive star.
For comparison, we also show the predicted yields from two rapidly rotating massive star models with v$_{\rm rot}=\, 300$ \kms\,\cite{Limongi2018}, which also give good $\chi^2$ fits to the data. These models include Pop II CCSN yields at [Fe/H]$=-2$ and $-3$, with additional contributions from weak s-process nucleosynthesis in their stellar winds. 
The rapidly rotating massive star models find similar predicted abundances for the elements that could be measured from optical spectra of the stars in C-19.  They also end as Wolf-Rayet stars (massive stars that have undergone significant mass loss), which have small ejecta masses that can create localized regions of chemical enrichment with inhomogeneous mixing.
We note that none of these models require carbon enrichment, which matches the normal carbon abundances we measure in C-19.
In summary, typical Pop II or Pop III massive stars can reproduce the average abundances in C-19 through rapid star formation in the early universe.


While the chemistry of C-19 suggests formation in a primordial burst of quick and intense star formation, in line with the properties of the first stellar generation in a globular cluster, 
its dynamically hot nature and global velocity dispersion $\sigma$(v$_{\rm los}$)$ \sim10$ km/s \cite{Yuan2025} are more consistent with a
stream from a dwarf-galaxy progenitor. This velocity dispersion is much higher than would be expected from the stream of a typical, baryon-dominated cluster \cite{TLi2022}, and several options to exist to explain this, e.g., C-19's progenitor 
may have been disrupted inside a low-mass halo before being disrupted by the Milky Way’s tidal forces \cite{Errani2022}, or the stream may have interacted with DM substructure in the MW halo \cite{Carlberg2025}.
Recent simulations have also suggested that systems like C-19 could be a natural product of early star formation in the standard cosmology and some can survive to the present day \cite{Taylor2025}.
The chemical abundance pattern and extremely low metallicity reflects enrichment processes driven by massive stars on very short timescales, i.e., in the first 100 Myr of cluster formation, at extremely early times in the evolution of the universe \citep{Bastian2018,
Yong2021Nature}. In these pristine environments, star formation efficiency can be extremely low \cite{Scannapieco2003,  Hirano2023} and the initial mass function top-heavy \cite{Wise2012, Chon2024}. When clusters form under these conditions, strong feedback from massive stars and their supernovae will drive rapid gas expulsion, leading to the gravitational unbinding of the cluster \cite{Baumgardt2007,Hirano2023}. 
N-body simulations of the tidal disruption of nearly unbound progenitors are able to match the observed dynamical structure of C-19 (see Methods).
Of course, star cluster formation at z=15 is currently not well understood, but C-19 may be giving us clues that such systems can form dynamically hot streams that survive to this day.

As fossils of early star formation like C-19 are now being discovered around the Milky Way, observations of the very distant universe with JWST raise the possibility of detecting such systems as they form.
Recent discoveries of extremely luminous objects at very high redshifts with JWST/NIRSpec 
have revealed elevated abundances of nitrogen \cite{Bunker2023,Cameron2023}.
%
%
While several explanations have been proposed, very early star formation and yields from Pop III CCSN and/or rapidly rotating massive stars are favoured \cite{Curti2025, Kobayashi2024}.
The significant difference in the predicted yields of N between these models can be seen in Fig.~\ref{fig:StarFit4}. Unfortunately, the optical spectra currently available for C-19 stars are insensitive to N.  However, the elements that we can study in the C-19 stars
(and cannot be measured in high-redshift systems) point us to the same interpretation: C-19 is the remnant of a primordial and rapid burst of star formation.
A similar star formation event in the early universe would be observable with JWST, e.g., LAT1-B at z\,=\,6.6 with M\,$<2700$\,M$_\odot$ \cite{Nakajima2025Nature}.
Thus, the C-19 stream provides us with a nearby glimpse of star formation events that occurred in the very early Universe and offers a unique opportunity to study the details of early star-formation, complementary to the enticing views provided by JWST of similar events in the high redshift universe.


%

%

\backmatter

\bmhead{Supplementary information}

If your article has accompanying supplementary file/s please state so here:
We will include a manuscript that details the spectral data acquisition, reduction, and full chemical analysis methodologies -- to be submitted to another journal \cite{Venn2025}.  The Discussion in that paper is under construction pending the acceptance of this Nature paper.



\bmhead{Acknowledgements}

Acknowledgements TBD


\clearpage

\begin{appendices}

\section{Extended Data}\label{secA}


\subsection{Observations \& Chemical Analysis}\label{secA1}

The Gemini High Resolution Optical Spectrograph
(GHOST) is a new fiber-fed bench-mounted spectrograph at the Gemini
South telescope that provides simultaneous wavelength
coverage with optimal performance from 363 – 950 nm \cite{McC2024GHOST}.
C-19 candidates were selected from two complementary stream-searching algorithms, STREAMFINDER \cite{Ibata2021} and StarGO \cite{Yuan2018}
combined with the low-metallicity samples from the Pristine Survey and Gaia BP/RP spectro-photometric catalogues \cite{Martin2024}. Candidates were confirmed from velocity and metallicity measurements by \cite{Yuan2022, Yuan2025}
; the best of these were selected for GHOST
commissioning, science verification, and follow-up programs (between February 2022 and September 2024), and one C-19 member in the north was observed at Subaru with HDS (Sept 2022). There is one candidate (C19F) that has consistent velocity from \cite{Yuan2025}, but no confirmed metallicity measurement, and is newly confirmed in this GHOST program. We further solidify C19N's membership by revealing that its abundances are highly consistent with those of other members, despite being the only one spatially separated by the MW disk. Combined with the published analyses of 4 stars taken with the GRACES spectrograph at Gemini North \cite{MartinVenn2022, Jeong2023} and one more from the Subaru HDS spectrograph \cite{Aoki2013}, we present Gaia IDs, coordinates, and radial velocities for 14 members of the C-19 stream in Table~\ref{tab:targets}.


Chemical abundances are determined from a classical model atmospheres analysis of the spectral features in each star. MARCS model atmospheres \cite{Gustafsson2008} were adopted for use with the 1D LTE radiative transfer code MOOG \cite{Sneden1973, Sobeck2011}.  Stellar parameters for extremely metal-poor stars (EMP; [Fe/H]$<-3$) are notoriously challenging.  Stellar parameters were determined using Gaia DR3 data, following the procedure in \cite{Sestito2023}.  Photometric effective temperatures are found using the colour-temperature calibrations from Gaia photometry \cite{Mucciarelli2020}, and surface gravities were determined using the Stefan-Boltzmann equation \cite{Bonifacio2025}.

Over 100 spectral lines of iron (mostly FeI), and nearly the same for all the other elements combined, were examined; line list and atomic data available in \cite{Venn2025}.
Corrections due to NLTE effects in the line formation were applied from the literature; Mg \cite{Bergemann2017}, Na \cite{Lind2012}, Al \cite{Lind2022}, Ca \cite{Mashonkina2017}, Fe \cite{Bergemann2012}, Sr and Ba \cite{Mashonkina2019}. 
Microturbulence values that removed any slope in the FeI line abundances with equivalent width were adopted examining both the LTE and NLTE FeI line abundances.  
Small abundance offsets ($>0.1$ dex) were found for some elements in the GHOST analysis of the standard star HD122563 when using the same line list used for the C-19 stars \cite{Venn2025} compared to a full line list analysis of HD122563 (which is bright, such that a much higher SNR spectrum is available providing many more weak lines \cite{Roederer2014}); these are noted in Table~\ref{tab:params}.

\subsection{Dispersions in elemental abundances}\label{secA2}
\label{sec:dispersions}
\begin{figure}[]
\centering
\includegraphics[width=0.7\textwidth]{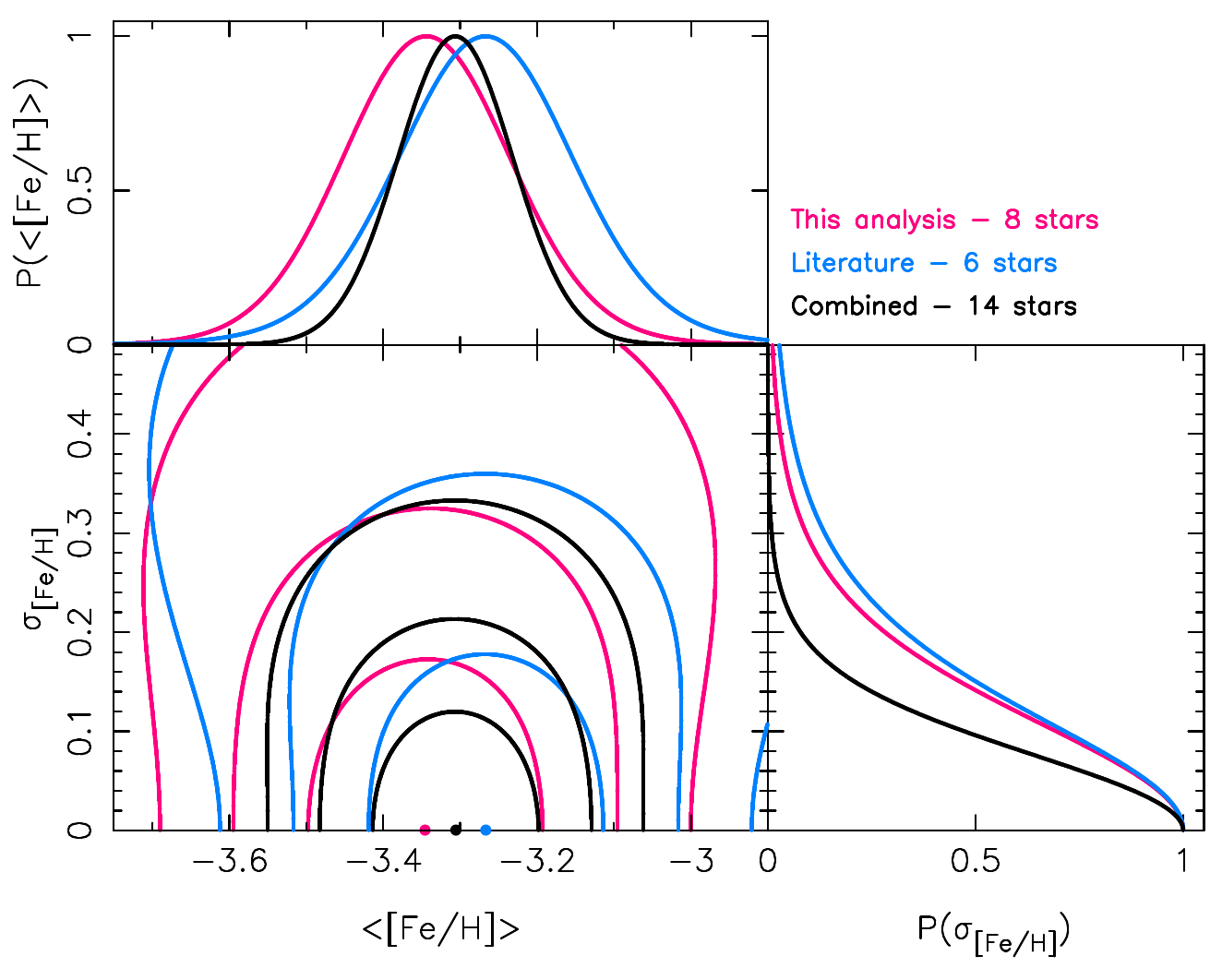}
\caption{Bottom-left panel: Two-dimensional likelihood function of the mean and dispersion of the [Fe/H] metallicity for C-19 based on the sample of 8 new stars presented here (magneta contours), on the 6 literature stars listed in Table~\ref{tab:params} (blue contours), and for the combined data sets that are independent (thick black contours). A dot represents the favoured model in each case. Top and right-hand panels: One-dimensional marginalized likelihood function for the mean and dispersion, respectively. From the combined sample of 14 C-19 stars, we infer that C-19 has a mean metallicity $\langle[\textrm{Fe}/\textrm{H}]\rangle=-3.31^{+0.07}_{-0.08}$ and a  metallicity dispersion $\sigma_{\mathrm{[Fe/H]}}<0.18$\,dex at the 95\% confidence level.
}\label{fig:FeH_contours}
\end{figure}

To infer the mean iron metallicity of C-19 stars and corresponding dispersion, we use a Gaussian likelihood model with two parameters, the mean, $\langle$[Fe/H]$\rangle$, and the dispersion, $\sigma_{\mathrm{[Fe/H]}}$. The likelihood is evaluated on a fine grid that covers possible values for these parameters and the resulting likelihood contours are represented in the bottom-left panel of Figure~\ref{fig:FeH_contours}. Marginalized one-dimensional likelihood functions are shown in the top and right-hand panels and allow us to conclude that the mean metallicity of the 8 new C-19 stars is $\langle$[Fe/H]$\rangle=-3.35\pm0.11$, with an unresolved metallicity dispersion 
$\sigma_{\mathrm{[Fe/H]}}<0.29$\,dex
at the 95\% confidence level.
This analysis is repeated for the 6 stars from the literature alone yielding $\langle$[Fe/H]$\rangle=-3.27\pm0.11$ and $\sigma_{\mathrm{[Fe/H]}}<0.33$\,dex at the 95\% confidence level (blue contours and lines in the Figure). Finally, using the combined sample of 14 stars (8 GHOST stars and 6 literature stars) provides our final metallicity inference for C-19: $\langle$[Fe/H]$\rangle=-3.31^{+0.07}_{-0.08}$ and $\sigma_{\mathrm{[Fe/H]}}<0.18$\,dex at the 95\% confidence level  (thick black contours and lines in the Figure).

\begin{figure}[]
\centering
\includegraphics[width=0.9\textwidth]{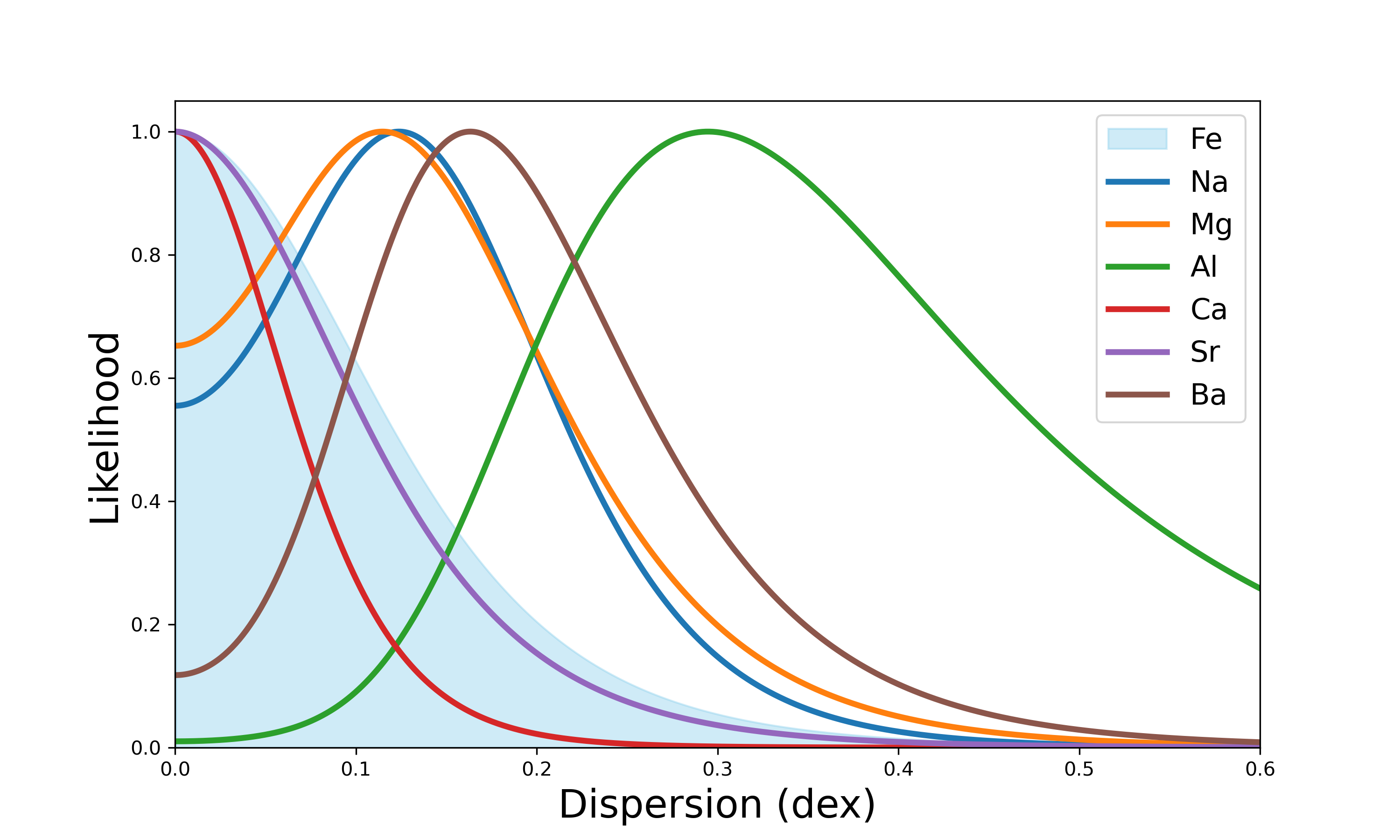}
\caption{
Likelihood functions of the Gaussian dispersion of 7 elements (Fe, Na, Mg, Al, Ca, Sr, and Ba) in the C-19 stars, available in Table~\ref{tab:params}.
}\label{fig:dispersions}
\end{figure}

The results of a similar analysis for different element abundances (Al, Ba, Ca, Mg, Na, Sr), still assuming a Gaussian likelihood model, are shown in Figure~\ref{fig:dispersions}. Ca and Sr, along with Fe, show no spread, with a dispersion that is consistent with zero and (very) small at high confidence. On the other hand, Mg and Na reveal a marginally resolved dispersion that nevertheless remains small, and Ba and Al have significant dispersions.

%

\begin{table}
\tiny
\caption{Target information for the 14 members of C-19 with detailed chemical abundances. Radial velocities (RV) in km/s.
}
\label{tab:targets}
\begin{tabular}{lccccccc}
\toprule
Name      & Gaia DR3 sourceID   & G & RA & DEC & INST & RV & REF \\
\midrule
C19A  & 2605574384366803968 & 15.88 & 346.2032242 & $-$10.6965325 & GHOST & $-132.8\pm0.2$ & \cite{Venn2025} \\
C19B  & 6559328209695612544 & 13.74 & 327.6864168 & $-$49.3800051 & GHOST & \, $+35.4\pm0.1$ & \cite{Venn2025} \\
C19C  & 6600784780223506944 & 13.47 & 339.9024531 & $-$32.7936701  & GHOST &  \, $-41.3\pm0.1$ & \cite{Venn2025} \\
C19D  & 6594796290142997376 & 14.51 & 335.4728668 & $-$38.3530749 & GHOST & \, $-19.6\pm0.2$ & \cite{Venn2025} \\
C19E  & 2641204161744171392 & 14.85 & 353.4703215 & $-$00.9700050  & GHOST & $ -134.2\pm0.2$ & \cite{Venn2025} \\
C19F  & 2658115921889849472 & 17.09 & 350.1429640 & +02.5838145 & GHOST & $-166.4\pm0.3$ & \cite{Venn2025} \\
C19G  & 2760807387346283648 & 16.39 & 351.3102187 & +07.9633947 & GHOST & $-178.4\pm0.1$ & \cite{Venn2025} \\
C19N  & 2288313499629002624 & 14.08 & 298.7454500 & +74.6564189 & HDS & $-134.3\pm0.7$ & \cite{Venn2025} \\
C19J  & 2865251577418971392 & 14.19 & 355.3224059	& +27.5993569 & UVES & $-183.6\pm0.3$  & \cite{Yuan2022} \\ 
C19K  & 2868052548930201984 & 15.69 & 354.7701575 & +30.2509843 & GRACES & $-186.7\pm2.2$ & \cite{MartinVenn2022} \\ 
C19L  & 2865368434887899008 & 15.39 & 355.1326831 & +27.9819596 & GRACES & $-194.4\pm2.0$ & \cite{MartinVenn2022} \\ 
C19M  & 2865256628300500352 & 15.81 & 355.2755506 & +27.7483341 & GRACES & $-197.3\pm2.1$ & \cite{MartinVenn2022} \\ 
C19W  & 2758373652717936640 & 14.65 & 354.5731253 & +09.0353924 & HDS & $-156.8\pm1.7$ & \cite{Aoki2013} \\ 
C19Y  & 2866151046649496832 & 14.28 & 354.9615142 & +28.4659616 & UVES & $-190.6\pm0.6$ & \cite{Yuan2022} \\
\botrule
\end{tabular}
\end{table}

\begin{table}
\tiny
\caption{Stellar parameters and chemistry for C-19 members.  The reported [Fe/H]$_{\rm NLTE}$ are the unweighted average of FeI and FeII (both NLTE).  Stellar parameter uncertainties (typically $\Delta$T$_{\rm eff}=\pm100$ K, $\Delta$log\,$g=\pm0.2$, $\Delta v_r\pm=0.2$, $\Delta$[Fe/H]$=\pm0.2$) are incorporated into the abundance errors in quadrature with the line measurement errors.  Spectral line lists and model atmosphere details in \cite{Venn2025}. 
}
\label{tab:params}
\setlength{\tabcolsep}{2.5pt}
\begin{tabular}{llccccccc}
\toprule
Target &  T$_{\rm eff}$/log\,$g$/$v_t$ & [Fe/H] 
 & [Na/Fe] & [Mg/Fe] & [Al/Fe] & [Ca/Fe] & [Sr/Fe] & [Ba/Fe] \\  
 & {\hspace{1.5cm}} & NLTE & NLTE    &  NLTE   &  NLTE   &   NLTE  &  NLTE  & NLTE \\
\midrule
C19A & 5055/2.5/1.5 & $-3.31\pm0.26$ & $-0.16\pm0.17$ & $+0.26\pm0.44$ & $-0.19\pm0.52$ & $+0.48\pm0.28$ & $-0.47\pm0.29$ & $-0.63\pm0.14$ \\
C19B & 4527/1.4/2.0 & $-3.37\pm0.30$ & $+0.40\pm0.18$ & $+0.25\pm0.21$ & $-0.94\pm0.17$ & $+0.42\pm0.25$ & $-0.17\pm0.34$ & $-0.70\pm0.31$ \\
C19C & 4436/1.0/2.5 & $-3.42\pm0.24$ & $-0.06\pm0.12$ & $+0.80\pm0.38$ & $-0.36\pm0.35$ & $+0.63\pm0.24$ & $-0.44\pm0.36$ & $-0.67\pm0.28$ \\
C19D & 4754/1.7/2.3 & $-3.17\pm0.32$ & ...            & $+0.52\pm0.41$ & $-0.61\pm0.17$ & $+0.53\pm0.20$ & $-0.73\pm0.31$ & $-0.71\pm0.29$ \\
C19E & 4691/1.3/2.2 & $-3.48\pm0.32$ & $+0.09\pm0.10$ & $+0.51\pm0.32$ & $-0.43\pm0.61$ & $+0.50\pm0.30$ & $-0.54\pm0.13$ & $-0.78\pm0.09$ \\
C19F & 5176/2.8/2.8 & $-3.26\pm0.31$ & $-0.23\pm0.16$ & $-0.03\pm0.14$ & $-0.16\pm0.13$ & $+0.58\pm0.16$ & $-0.40\pm0.18$ & $-0.30\pm0.13$  \\
C19G & 5109/2.1/2.0 & $-3.26\pm0.34$ & $-0.05\pm0.11$ & $+0.09\pm0.14$ & $-0.11\pm0.14$ & $+0.45\pm0.14$ & $-0.45\pm0.21$ & $-0.64\pm0.40$ \\
C19N & 4647/1.5/2.9 & $-3.40\pm0.24$ & $-0.01\pm0.29$ & $+0.23\pm0.40$ &  ...           & $+0.39\pm0.24$ & $-0.42\pm0.14$ & $-0.84\pm0.32$ \\
C19J & 4569/1.0/2.2 & $-3.46\pm0.33$  & $+0.20\pm0.28$ & ...            & $+0.17\pm0.34$ & $+0.62\pm0.21$ & $-0.29\pm0.28$ & $-0.69\pm0.20 $ \\
C19K & 4928/1.8/2.1 & $-3.23\pm0.21$  & $+0.34\pm0.18$ & $+0.33\pm0.17$ &  ...           & $+0.56\pm0.20$ & ...            & ...            \\
C19L & 4881/1.6/2.2 & $-3.29\pm0.20$  & $+0.37\pm0.17$ & $+0.39\pm0.14$ &  ...           & $+0.56\pm0.08$ & ...            & $-0.22\pm0.17$ \\
C19M & 4958/1.9/2.1 & $-3.20\pm0.24$  & $+0.05\pm0.17$ & $+0.42\pm0.14$ &  ...           & $+0.48\pm0.13$ & ...            &  ...           \\
C19W & 4900/1.9/1.5 & $-3.10\pm0.50$  & ...            & ...            &  ...           & ...            & ...            &  ...           \\
C19Y & 4446/0.9/2.2 & $-3.28\pm0.22$  & $+0.19\pm0.24$ & ...            & ...            & ...            &  ...           &  ...           \\
\textit{Offset}*  & ... & ... &  ... & $-0.2$ & ... & ... & ... & $-0.3$ \\
\botrule
\end{tabular}
\begin{tablenotes}
\footnotesize
\item[*]\textit{Offset} refers to an abundance offset ($>0.1$ dex) found for some elements in the analysis of the standard star HD122563 when using the same line list used for the C-19 stars \cite{Venn2025} compared to a full line list analysis of HD122563 \cite{Roederer2014}).
\end{tablenotes}
\end{table}

\subsection{Comparison Stars and References}\label{secA3}

The Figures in this paper have included chemical abundances from decades of published work.
Our Milky Way field star abundances are taken from two main references: (1) those collected in the SAGA database \cite[][ with updates in 2021]{Suda2008}, which include measurements for metal-poor MW field stars from \cite{Aoki2013, Yong2013, Yong2021,
Roederer2014}, and (2) a recent and homogeneous analysis using Subaru HDS spectra \cite{HLi2022}.
Chemical abundances for two very metal-poor globular clusters (where [Fe/H] $<-2.5$) are from the literature: M15 \cite{Sobeck11, Meszaros2020} and M92 \cite{Meszaros2020, Kirby2023}.
Similarly, the chemistry for three very low-mass globular clusters (M $<10^5$ M$_\odot$ are from the literature: Ter8 \cite{Carretta2014}, IC4499 \cite{Dalessandro2018}, NGC6535 \cite{Bragaglia2017}.
%

Element abundances for the UFD galaxies are initially from the SAGA database \cite{Suda2017}, which includes the following measurements:
Bootes I \citep{Feltzing09,  Norris10, Gilmore13, Ishigaki14}, 
Bootes II \citep{Ji16},
Carina II \citep{Ji20}, Carina III \citep{Ji20}, Coma Berenices \citep{Frebel10, Vargas13}, Grus I \citep{Ji2019}, Grus II \citep{Hansen2020}, Hercules \citep{Koch08, Aden11, Vargas13,  Francois16}, Horologium I \citep{Nagasawa_2018}, Leo IV \citep{Simon2010, Francois16, Vargas13}, Pisces II \citep{Spite2018}, Reticulum II \citep{Ji16}, Segue 1 \citep{Norris10, Frebel14}, Segue 2 \citep{RoedererKirby2014}, Triangulum II \citep{Venn2017, Kirby_2017, Ji2019}, Tucana II \citep{Ji16, Chiti18}, Tucana III \citep{Hansen2017, Marshall_2019}, Ursa Major II \citep{Frebel10}.
We also update with new data for some systems:
Bootes I \citep{Waller2023},
Cetus II \citep{Webber2023}, Coma Berenices \citep{Waller2023}, 
Reticulum II \citep{Hayes2023}, 
Tucana II \citep{Chiti_2023}, 
Tucana V \citep{Hansen2024}, Ursa Major I \citep{Waller2023}, and
updated NLTE abundances for Com Ber from \cite{Sitnova2021}.
Finally, we add the newest chemical abundances for UFDs from \cite{Dovgal2025}, which includes new Gemini GHOST spectral analyses of 1-2 stars in each of Boo2, RetII, Eri4, Tuc4, Col1, Phx2, Del2, Pic1, and Pic2.

\subsection{Simulations} \label{secA4}

We explore a scenario that the C-19 progenitor cluster has undergone rapid gas expulsion using the N-body code \texttt{PeTar} \cite{Wang2020a}. While the cluster is in an expansion phase, it begins its orbit around the Milky Way, starting from apo-center at $\sim$24 kpc from the Galactic center, which is traced back from the current stream orbit \cite{Ibata2024}.
After evolving in the Galaxy for $\sim10$ Gyr, the cluster is disrupted and leaves star particles along its current orbit.

A similar framework has been used to simulate the Palomar 5 stream \cite{Wang2023} except for the Milky Way potential from \cite{McMillan2017} with a total viral mass of $\sim1.30\times10^{12}$M$_{\odot}$ used in this work. The cluster mass is 10$^4$M$_{\odot}$ modeled by 160,000 star particles in the half-mass radius of 1 pc. We adopt a universal initial mass function (IMF) from \cite{Kroupa2001}. Initially, the cluster is set to be in a rapid expansion phase after gas expulsion. This is implemented by inflating the velocities of stars under virial equilibrium state by a factor of 4.0. In this scenario, the total mass of gas before the expulsion is 16 times of the stellar mass afterwards. Assuming that the gas removal happens instantaneously, the star formation efficiency, \ie fraction of gas that is converted to stars is 1/16 or $\sim$0.06 \cite{Baumgardt2007}. The very low star formation efficiency is expected for Pop III clusters formed in metal-free environments \cite{Scannapieco2003, Hirano2023, Chon2024}.

Fig.~\ref{fig:orbit} shows that the simulated particles in this simple scenario form a structure that resembles the C-19 stream, with similar width on the sky and velocity dispersion ($\sim10$\,km/s). Future simulation studies by Wang et al. (2026, in prep) will explore the impacts on the simulated stream from variations of IMF, including a top-heavy IMF \cite{Marks2012} and variations of the initial binary and black hole fractions in the cluster \cite{Gieles2021, Wang2023}, though preliminary results show that the initial rapid expansion at early times is the key for reproducing the C-19 stream properties.


%

\end{appendices}


\bibliography{C19-bibliography}

\end{document}